# Away from resolution, assessing the information content of super-resolution images

**Thomas Pengo[1,2], Nicolas Olivier[1,3], Suliana Manley[1]**

1 Laboratory for Experimental Biophysics, School of Basic Sciences, Swiss Federal Institute of Technology (EPFL), Lausanne, Switzerland
2 Advanced Light Microscopy Unit and Design of Biological Systems, Center for Genomic Regulation, Barcelona, Spain
3 Department of Physics, King's College London, UK

**Abstract**

**Super-resolution microscopy has revolutionized optical fluorescence imaging by improving 3D resolution by 1-2 orders of magnitude. While different methods can successfully increase the resolution, all methods share significant differences with standard imaging methods, making the usual measures of resolution inapplicable. In particular image quality and information content are spatially heterogeneous with variabilities that can be comparable to their mean values, limiting the use of the average resolution as a predictor for local information. A common use of super-resolution data is to test or establish structural models, and in these cases it would be valuable to assess the capacity of the data to validate a model. We focus here on single-molecule localization methods and present a new way of assessing the quality and reliability of super-resolution data.**

## Introduction

Super-resolution fluorescence imaging offers an exciting glimpse into the organization of specific biological structures below the diffraction limit. Improvements in methods, analysis, and dyes have aided the commercialization and dissemination of super-resolution imaging in the form of stimulated emission depletion (STED) microscopy (1), structured illumination microscopy (SIM) (2) and single molecule localization microscopy (photoactivated localization microscopy (PALM)/stochastic optical reconstruction microscopy (STORM)) (3-5). An important aspect of properly applying these methods and interpreting their results is to accurately assess the extent to which one can infer information about the sample based on the acquired data. Image quality is commonly characterized by a single value, resolution. One approach to estimate resolution is by empirically measuring the cross-sectional profile of features of known size that are resolution limited (6-8). Since this takes as input a select portion of the image, it is an inherently local measure, and the extension of this value as the image resolution implicitly assumes a constant resolution over the whole image. An alternative approach is to calculate a global theoretical estimate for resolution (9-13), including one based on Fourier correlation of images (14). However, image quality and information content are spatially heterogeneous, so the average global value is of little use when estimating a local feature. Since a common use of super-resolution data is to test or establish structural models, we focus here on these cases and go beyond the resolution, instead assessing the capacity of the data to validate a model.

Here, we focus primarily on single molecule localization microscopy (referred to as LM) imaging, where the main experimental determinants of spatial resolution are the density of localized molecules on the structure of interest, together with their localization precision. In this class of super-resolution imaging, an almost arbitrary localization precision can be set by removing poorly localized molecules, but there is a trade-off since the molecule density is concomitantly reduced (3). In contrast with diffraction limited microscopy, in which resolution is typically sufficient to describe the quality of the imaging system, here, we need at least two interdependent values (brightness and density of molecules) to characterize the image. Moreover, because both values can vary greatly across an image, they are not sufficient to assess the ability to measure a given feature. We propose and demonstrate an alternative metric for image quality: the bootstrapped confidence in a given measurement (BMC). This avoids many of the drawbacks of current approaches, while still setting limits for the interpretation of data. We further introduce a new software tool to facilitate averaging over image features to reduce the effect of heterogeneity and increase the statistical significance of our measurements.

## Materials and Methods

Optical setup: STORM Imaging was performed according to (21). Briefly, we used a modified Olympus IX71 inverted microscope with a 641 nm laser (Coherent, CUBE 640-100C, USA) reflected by a multiband dichroic (89100 bs, Chroma,USA) on the back aperture of a 100x1.3NA oil objective (Olympus, UplanFL, Japan) mounted on a piezo objective scanner (P-725 PIFOC, Physik Instrumente, Germany). The collected fluorescence was filtered using a band-pass emission filter (ET700/75, Chroma) and imaged onto an EMCCD camera (IxonEM+, Andor) with a 100 nm pixel and using the conventional CCD amplifier at a frame rate of 25 fps. Laser intensity on the sample measured after the objective was 2–5 kW.cm$^{-2}$ (typically 40 mW in a 200 μm$^2$ region and 10,000–20,000 frames were recorded.

SIM imaging (Figure 1) was performed on a N-SIM microscope (Nikon, Japan) STED imaging (Figure 1) was performed on TCS STED CW microscope (Leica, Germany)

Sample preparation: COS-7 cells were cultured in DMEM supplemented with 10% FBS (Sigma-Aldrich) in a cell culture incubator (37C and 5% CO2). Cells were on cleaned 25 #1 coverglass (Menzel). 24 h after plating, cells were pre-extracted for 20 s in 0.5% Triton X-100 (Triton) in BRB80 (80 mM PIPES, 1 mM MgCl2, 1 mM EGTA, adjusted to pH 6.8 with KOH) supplemented with 4 mM EGTA, washed in PBS, fixed for 10 min in -20C Methanol (Sigma-Aldrich), then washed again in PBS. The samples were then blocked for 30 minutes in 5% BSA, before being incubated for 1.5 h at room temperature with 1:1000 mouse alpha-tubulin antibodies (Sigma,T5168) in 1% BSA diluted in PBS 0.2% Triton (PBST), followed by 3 washes with PBST, and then incubated for 45 min in 1%BSA-PBST with 1:1000 goat anti-mouse Alexa Fluor 647 (Alexa-647) F(ab')2 secondary antibody fragments (Life Technologies, A-21237) or Alexa-647 Full length secondary antibody (Life technology A-21235). Alternatively, primary mouse anti-alpha-tubulin antibodies were directly conjugated using the APEX Alexa Fluor 647 antibody labeling kit (Life Technologies) according to the manufacturer's instructions. For Cep152 imaging, U2OS cells were maintained in McCoy's 5A GlutaMAX medium (Life Technologies) supplemented with 10% FBS and on cleaned 25 mm #1 cover-glass (Menzell). Fixation and immunostaining was performed similarly as for tubulin, except that the primary rabbit anti-Cep152 antibody (Sigma-Aldrich, HPA039408) was used at 1:2000 in 1% BSA - PBST, and the secondary antibody was goat anti-rabbit Alexa-647 F(ab')2 secondary antibody fragments (Life Technologies, A-21246) at 1:1000 in 1% BSA - PBST. For SIM and STED imaging, Alexa-488 F(ab')2 secondary antibody fragments (Life Technologies, A- 11017) were used instead.

Imaging Buffer: STORM imaging was performed according to (21) in 10 mM PBS-Tris pH 7.5 with 10 mM MEA 9 (MEA – Sigma-Aldrich 30070) combined with 50 mM β-mercaptoethanol (BME - Sigma-Aldrich M6250), 2 mM Cyclooctatetraene (COT – Sigma-Aldrich 138924), 2.5 mM PCA (Protocatechuic acid, Sigma-Aldrich 37580) and 50 nM (Protocatechuic dioxygenase, Sigma-Aldrich P8279) .

Data Analysis: Single molecule localization was performed using Peakselector (Courtesy of H.Hess). Outliers (peaks detected for more than 15 consecutive frames, and peaks not fitted properly with a Gaussian function) as well as peaks localized with less than 1000 photons were removed from the analysis. Peaks detected in successive frames at a distance of less than 60 nm were considered as originating from a single molecule and grouped. Analysis software for BMC calculations was written in MATLAB (The MathWorks, Natick, MA) as a plugin for PALMsiever (15).

**Results and Discussion**

Measuring a cross-sectional profile of a known, sub-diffraction limited structure has been one accepted way of measuring image resolution. It consists of selecting a region of interest (ROI) and determining its intensity profile. Microtubules, which are tubular cytoskeletal elements, are often used as a de facto benchmark for super-resolution techniques, since their outer diameter is known to be 25 nm with very small variability (16). Images of immunostained microtubules were taken with three major super-resolution techniques: STED, SIM, and LM (Fig. 1A). Three profiles are shown for each image, demonstrating that the assumption that different ROI from the same structure should show similar profiles does not always hold. Imaging conditions, spurious structures from staining, and photon noise can all influence the image quality and deteriorate the profiles, resulting in high variability from region to region.

This highlights the overall difficulty in applying cross-sectional profiles as a general measure of resolution, since it requires the selection of ROI(s). This choice can lead to values that are not representative of the overall quality of the image. Therefore, this resolution measure is not appropriate to determine whether another structure in the image is clearly resolved.

For LM, an alternative approach is to combine the localization density and precision in an adapted form of the Nyquist sampling criterion (17), to define a threshold where the density would be high enough to make the localization precision the effective resolution (10). However, this is an overly optimistic estimate; thus a more global measure of the resolution based on frequency analysis was proposed (12). A statistically and mathematically rigorous analysis was used to derive a relationship between sample structure, labeling density and localization precision. We applied this analysis to a sample with similar characteristics to microtubules, a Gaussian spatial frequency spectrum with a characteristic size of 35 nm and a localization precision of 10 nm (Fig. S1A). As predicted by the theory, for any chosen cutoff signal-to-noise ratio (SNR), the corresponding spatial cutoff frequency increases as the logarithm of the square root of the number of emitters (Fig. S1B) compared to an optimistic square root for the Nyquist threshold.

In imaging of most biological specimens, global measures of image quality may be misleading, due to the high degree of heterogeneity of information content within an image (Fig. 1A, Fig. 2B). We further illustrate the effects of such heterogeneity with synthetic data (Fig. 1C). Even though parts of the image show reduced information content (red) in which the structural frequencies of the sample cannot be recovered, other parts (green) contain sufficient information to distinguish the image features. Another global measure, based on FRC resolution (14) can also be shown to vary greatly when applied to sub-regions of the same image (Figure S2), illustrating the variability in the resolution. Moreover, global measures treat all portions of the image the same, regardless of whether they contain biological features of interest. We therefore propose to adapt the local measures, making them more robust by adding statistical tests to be able to judge the significance of the measure.

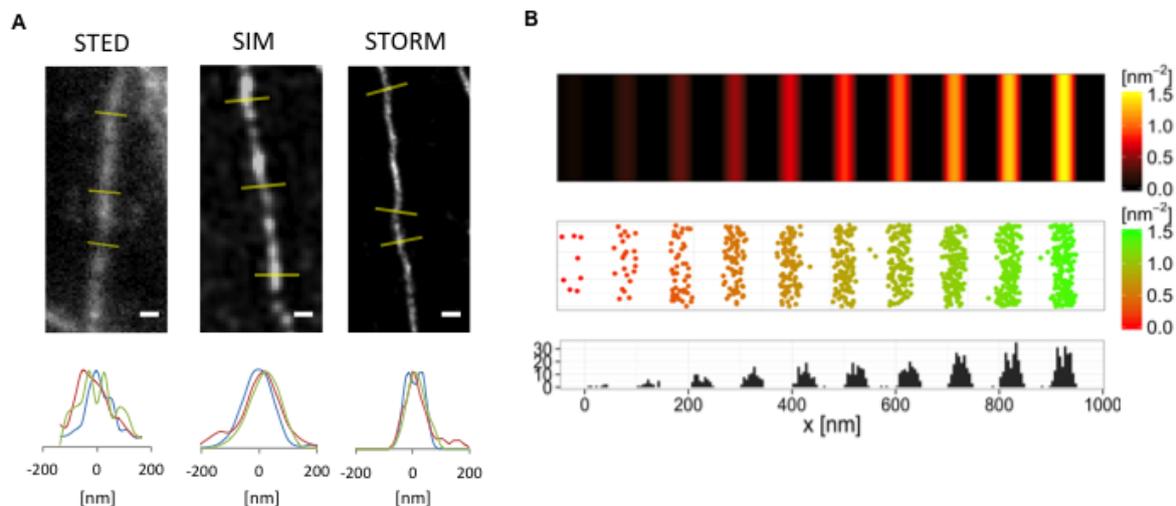

**Figure 1**. Two current measures of resolution. (A) Images and intensity profiles from STED microscopy, SIM and STORM. Scalebars 100 nm. Three profiles extracted from each of the images

(regions indicated by yellow lines) are shown below. (B) An example synthetic image where the average spectrum is a poor basis for quantifying resolution. Sampling density is a function of position (x) within the image.

We start by revisiting the problem of measuring resolution using microtubule images. Since the size and structure of microtubules are known (18) we can make some assumptions. First, the primary antibodies we are using are located at the outside of the tube according to their target epitope(19), and they are ~10 nm in size. Second, we assume their lateral profile should be constant along the tube. Therefore, we can determine an average profile by tracing the microtubule along its estimated centerline (Fig. 2A). Third, their 2D projection should yield a double peak (20-22).

We automated this analysis by implementing tracing software in which the user selects an initial seed point and direction and the algorithm steps along the direction of highest density until either no points are found or the edge of the window is reached. Given the trace, points can then be collected and expressed in terms of the trace coordinate system, effectively removing curvature from the microtubule structure. A histogram of the distance of each point to the centerline reveals the underlying profile of the structure. Without such a tool, one could take individual short segments of the trace (200 nm in the example shown), but this would give a significant variability (Fig. 2B). To estimate the variability, a single Gaussian was fitted to each section, giving a full width at half-maximum of 53 +/- 8.7 nm, while averaging over a 3.5 micron length brings the error down to an estimated 3.9 nm. This improvement is due to the effectively higher molecule density of the averaged data.

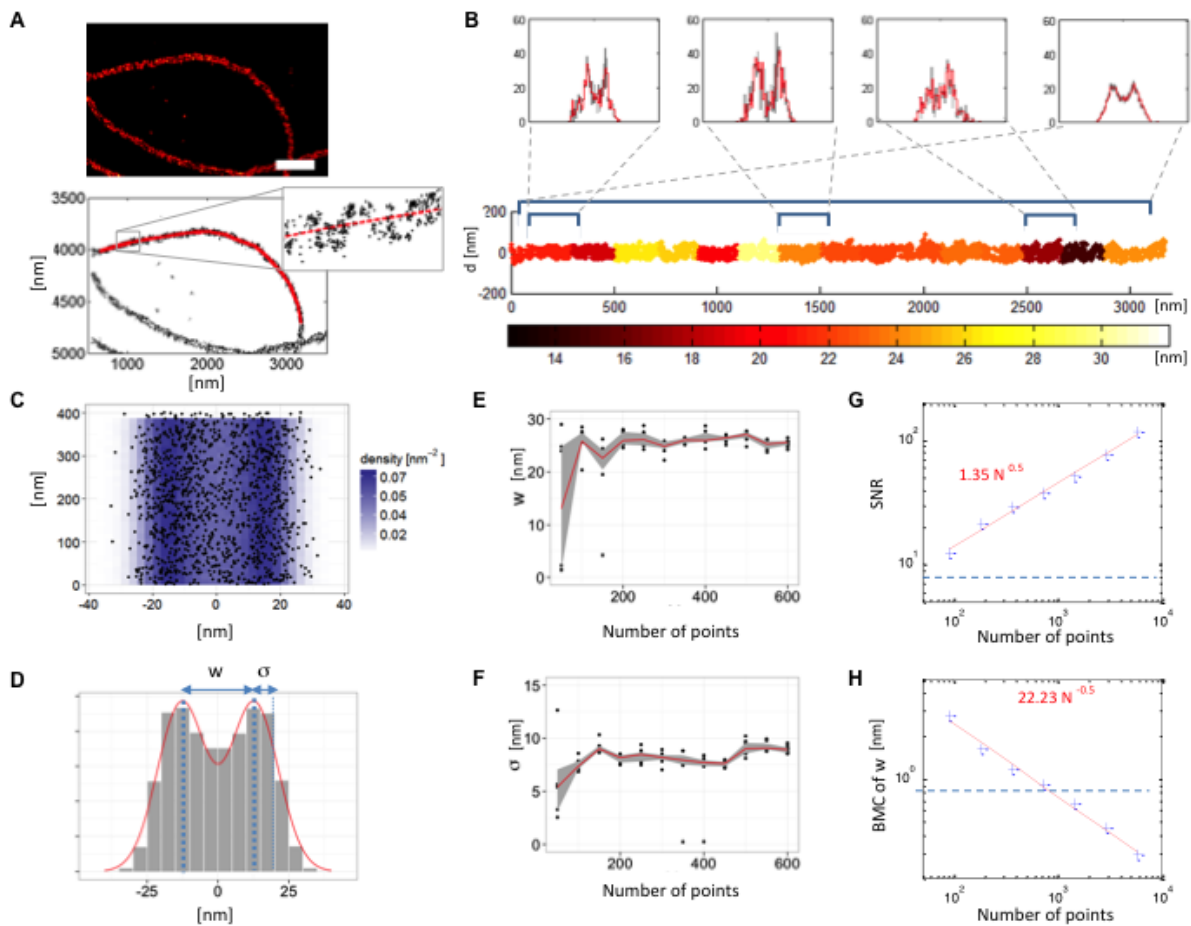

**Figure 2**. Solutions to both better estimate the quantity of interest and the error on the estimation. (A) Example of the tracing algorithm. Above: curved section of a microtubule (scale-bar 500nm). Below: same section showing individual localized molecules. From a seed point, the tracing algorithm steps along the direction of maximum alignment adding steps to the trace (red) until either an insufficient number of unvisited points is found, or the edge of the region is reached. (B) Visual representation of the variability of estimating the width of the microtubule for the same section in (A). Around each trace step, the detected peaks are binned into a histogram and fitted with a Gaussian function. The fitted standard deviation parameter is plotted (color bar). (C) Bootstrapping concept on a simulated section of microtubule. 1500 points from the simulated section of microtubule. (D) The histogram of the projection along the transversal axis, along with a fitted double Gaussian. (E-F) Convergence of the estimation of w and σ with increasing number of points. (G) SNR (mean over standard-deviation) of the estimation of w, depending on the number of points. (H) Estimated error (bootstrapped standard-deviation of w) as a function of the number of points. All units are nanometers unless otherwise stated.

Using this method, we mitigate one of the main problems of using line profiles to estimate the localization precision, which is the variability. It is however still not a measure of the resolution or image quality *per se*: the variability in localization precision is lost, only the mean value remains, and it does not take into account the density of molecules.

We propose to use bootstrapping to assess the confidence in a measurement and hence quantify the quality of an image for making the measurement. This can be applied in combination with tracing or particle averaging (23-25) (if there is *a priori* information about the sample that can be used) or without. Specifically, we define BMC (Bootstrapped Measurement Confidence) as the standard deviation of the bootstrapped resampling distribution on the measurements. For example, to find the BMC on the width of a structure such as microtubules (Fig 2C), one would fit a double Gaussian to the profile (Fig 2D) and estimate the variance on the fitted width. Bootstrapping repeatedly samples from a given population and repeatedly calculates the given statistic (in this case, a fit to the microtubule profile). At this point one can assess how the measurement parameters fluctuate and converge (Fig. 2E,F) or become more reliable (Fig. 2G,H) with an increasing number of localizations. With our definition of BMC, lower values correspond to lower variability and higher repeatability of the measured statistic. As one would expect, this decreases with increasing numbers of localized points in the image.

We applied this method to quantify the confidence on the width of microtubules. We used different labeling methods to change the effective tube size; either directly conjugated primary antibodies (CONJ), secondary antibody fragments (FAB), or full-length secondary antibodies (FL). We expected that the distance between the two peaks would increase as a function of the size of the antibodies. To perform the quantification, we select a trace from a rendered image. The histogram of the cross-section for a representative data set is shown to converge after a few thousand points (Fig. 3B). Fitting the cross-section from 1000 points to a double Gaussian gives a width of 34 nm with a BMC < 1 nm. Signal to noise increases and BMC decreases with the number of points following a power law with an exponent of around 0.5 (Fig. 3D,E), in agreement with the simulated data (Fig. 2G,H). Thus, for measuring the width of this structure, the BMC is effectively related to the square root of the number of points.

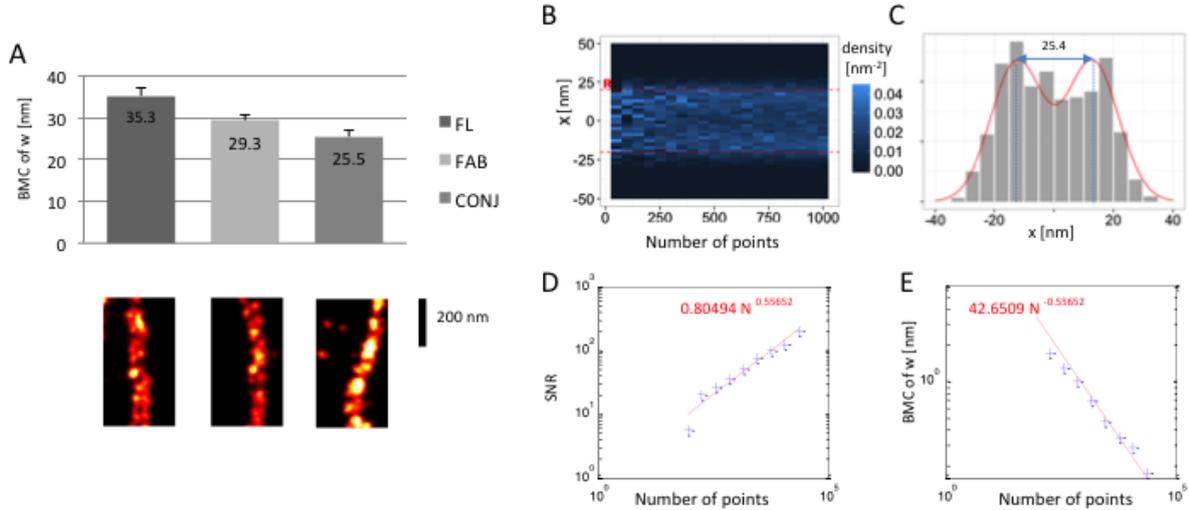

**Figure 3**. (A) Our algorithm to assess confidence was used to identify the difference in tube radii, depending on the labeling conditions. (B) Cumulative normalized histogram for a 8.5um microtubule stained with full length antibodies. The histogram shows how the profile of the microtubule becomes clearer with increasing number of detections. (C) Profile resulting from the tracing algorithm and the fitted double Gaussians. (D) SNR dependence on the number of points shown in log-scale. (E) BMC of the estimated width plotted against the number of points.

We used BMC to compare the significance of differences between the three different staining techniques (Fig. 3A). The BMC in the three cases, containing N=1854, 3821, and 1868 localizations is 2.0, 1.2 and 1.7 nm respectively, shown in the graph as an error bar. This estimate of the error in our experimental data makes sense. For example, if we had N estimations of the width and we averaged, we would expect an error (supposing a Gaussian distribution of the error) of sigma/sqrt(N), which in our case would give 35,3/sqrt(1854) = 0.82 for the full-length antibodies. Furthermore, the effective radius is consistent with the bare diameter of 25 nm broadened by either an isotropically labeled 10 nm structure, (CONJ), or a 10 nm primary antibody with 5 nm (FAB2) or 10 nm (FL) secondary antibodies respectively (Figure S4). According to Fitzgerald *et al.* (12), to attain resolutions of this order, we would need a density well in excess of 1 molecule/$nm^2$ (Figure S1), while our data has on the order of 0.05 molecule/$nm^2$.

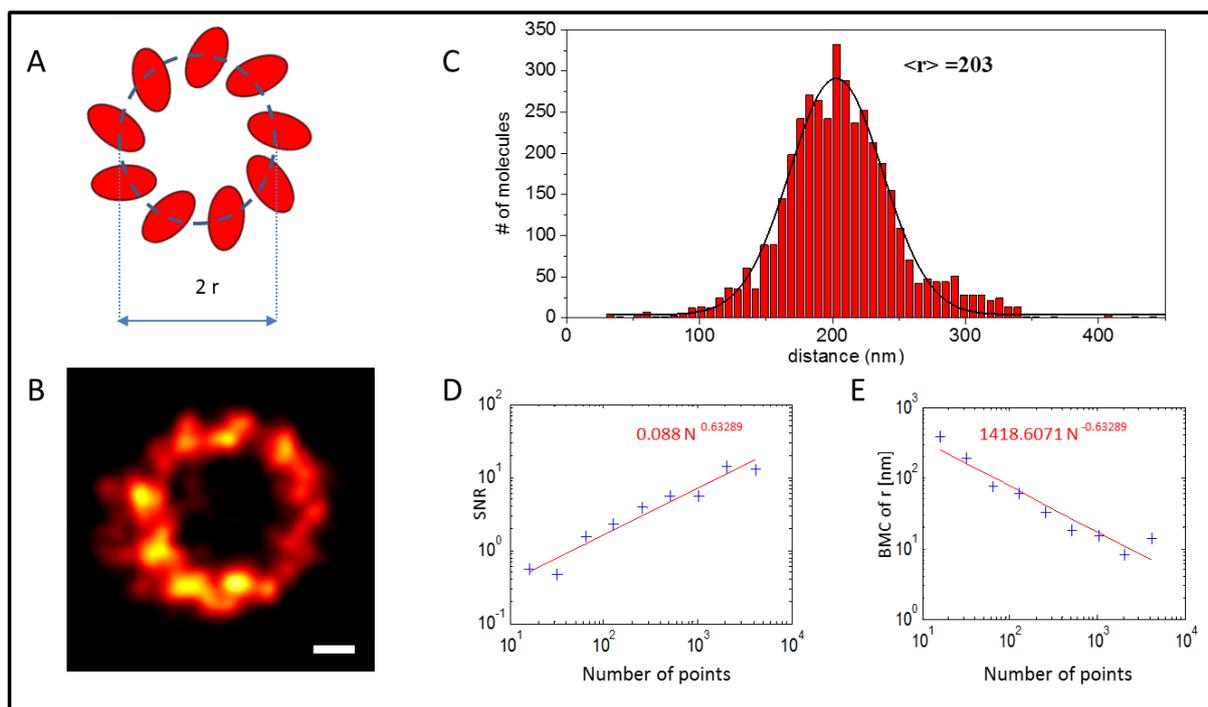

**Figure 4.** (A) Model of the centrosomal distribution of CEP 152, with the 9 sub-units and (B) super-resolution image (scale-bar 100 nm). (C) Radial profile, where the radius is estimated by fitting the distance distribution to a Gaussian. (D) Our algorithm assessed SNR for the data and (E) estimated the BMC on r.

We also applied our method to the measurement of the radius of the centriolar protein CEP152, predicted to be distributed along a tube and therefore projected along its long axis to form a ring (distribution schematized in Figure 4A (21,22)). A rendered image of a centriole stained with antibodies against CEP152 (Fig. 4B) was obtained under the same conditions described for the microtubules (Fig 2) (26). Figure 4C-E show the analysis of the radius: 4C shows the radial distribution of the points, fitted to a Gaussian distribution. The measured radius is 203 nm with an estimated bootstrapped confidence of 1 nm. The SNR and the confidence are plotted as a function of the number of points in Fig. 4D-E. Both exhibit a power-law behavior with an exponent of 0.63. This rate of convergence is higher than the rate found on microtubules, which makes sense since for the this structure we fit a radial profile using a single Gaussian with only three degrees of freedom: center (the radius of the radial profile), amplitude and sigma. On the other hand, the microtubule profiles are fit with a double Gaussian, which has four parameters: center, amplitude, sigma and distance between the peaks. We expect that the more degrees of freedom in the fitting function, the slower the convergence.

## Conclusion

Variability has always existed in optical microscopy: aberrations, and non-uniform signal to noise being the most common sources usually encountered in "classical" methods. However, with the advent of super-resolution microscopy methods, the variability in the (local) resolution is now comparable to the average resolution. In the case of Localization Microscopy, both the localization precision and the density have distributions with high variability that have the same order of magnitude as the averages. This makes using the average value as a criterion for local information inapplicable (as has recently been shown for cyo electron microscopy (27)), and requires new ways of estimating the reliability of a given measurement.

Our proposed measure of confidence provides a much-needed complement to current measures, which focus on mean resolution. Similar to intensity profiles and Fourier ring correlation, it is based on experimental data. Furthermore, it can be used to assess the robustness of specific measures of features of interest, thereby more directly evaluating the quality of the data with respect to interpretation of its content. This is an important aspect for biological imaging, where one would like to be able to comment on the sizes of features and estimate the reliability of their measured size. We focused here on LM, but other SR methods also suffer from high variability, which makes resolution an insufficient metric of image quality and the validity of measurements on features. Although BMC as applied here relies on the composite nature of LM it can also be applied with some slight modifications to assess the reliability of data from other SR methods. More precisely, BMC could be performed on repeated measurements, be it through iterative imaging of the same structure, multiple segments of continuous structures like microtubules, or multiple instances of the same structure within an image.

We confirmed the validity of the bootstrapping method (Supp. Fig. 5 and 6). We also tested its dependence on the number of points N and the localization precision. Both follow the Cramer-Rao bound closely, but slightly overestimate the error as one would expect. The estimated bootstrapped values on the error in our experimental data are also in line with expectation.

The analysis performed here showed good agreement in the dependence of SNR and confidence on the number of points between synthetic and real data from microtubule images. Both went as a power law, with powers of +/- 0.5. We also analyzed a different structure, the centriole, which has an entirely different symmetry, and found there power laws of +/- 0.6. For both structures, we found an intermediate dependence to the ones previously proposed, which were likely either too generous (10) or too strict (13). It is interesting to note that different structures give rise to different power laws, a feature that could be difficult to capture with a purely theoretical definition of resolution. This highlights the importance of experimental estimates of data quality.

**Acknowledgements**:

The authors thank C. Bauer at the University of Geneva for help with STED imaging, D. Keller and C. Ben Adiba for help with sample preparation, H. Hess for the use of Peakselector software, and J. Fitzgerald for discussion.

**Grants**:

This work was supported by a European Commission Seventh Framework Programme grant No. 243016 (PALMassembly) and the Brazilian-Swiss Joint Research Program, No. 011004.

**Supplementary Data**

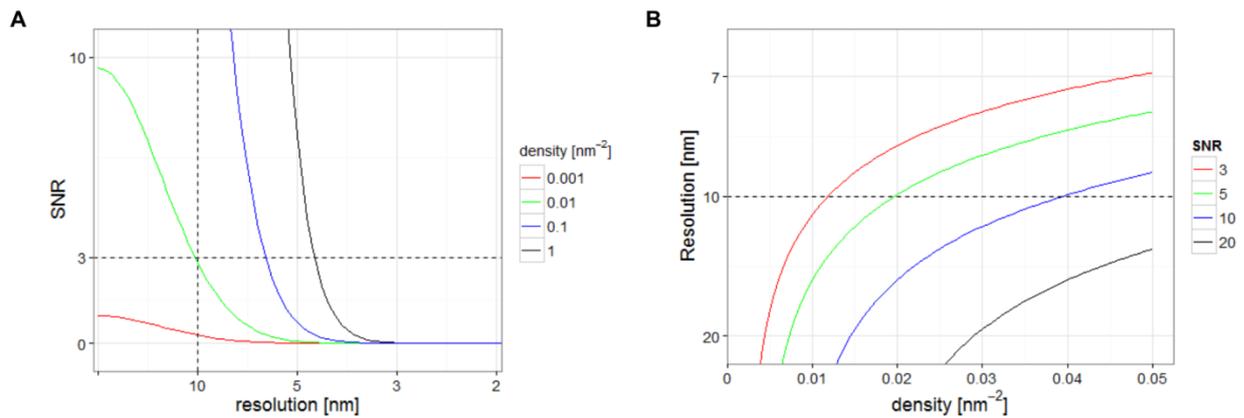

**Figure S1: Estimation-theoretical estimation of average resolution**

Estimation-theoretical average resolution **by** Fitzerald *et al.* based on a Gaussian-shaped **frequency** spectrum. **A**. Attainable signal-to-noise-ratio (SNR) for different spatial resolutions (here the inverse of spatial frequency) and spatial densities. **B**. Attainable resolutions with varying densities and SNRs. Notice how slowly increasing density improves resolution after the inflexion point.

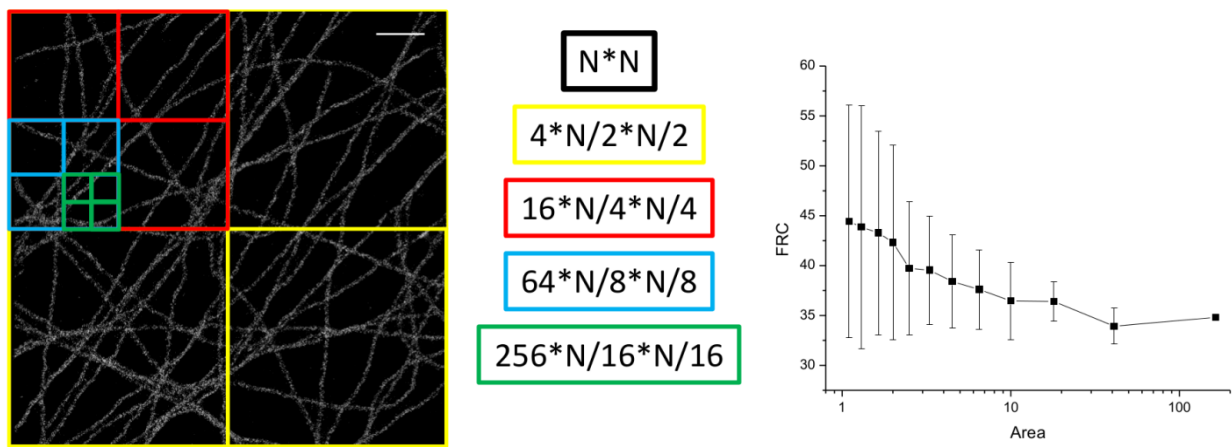

**Figure S2: Variability in local FRC resolution**

FRC resolution is computed on regions of decreasing area (from 128 to 1 μm$^2$) of a STORM image of microtubules, and the average value and standard deviation are plotted as a function of area.

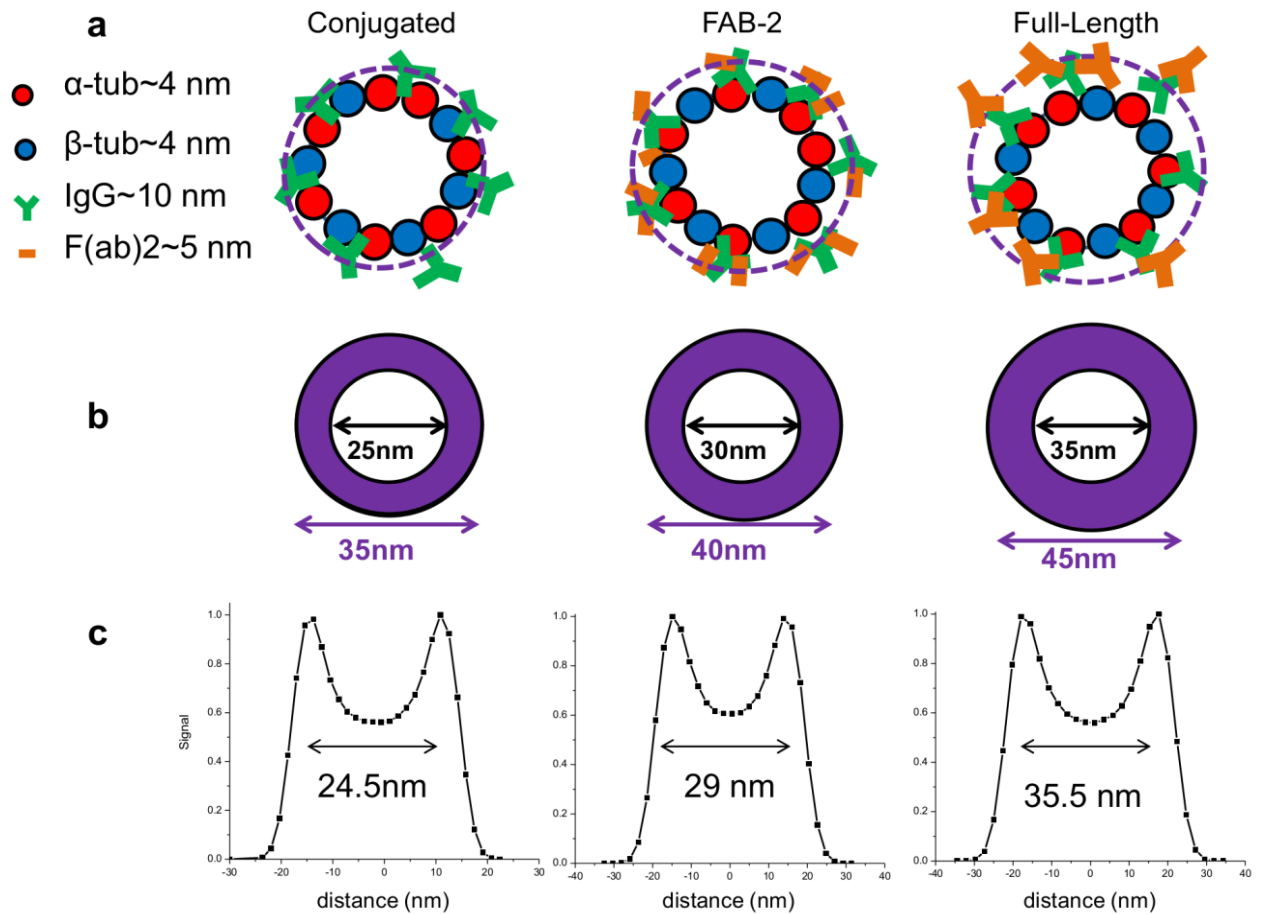

**Figure S3: Model of antibody-stained microtubules.**

(a) Schematics of microtubule structures with α-tubulin and β-tubulin subunits forming a 25 nm diameter tube. The antibodies used in this paper bind to the outer domain of α-tubulin, and depending on the staining used, the dyes are either attached directly to the primary antibody ("Conjugated"), to a secondary antibody fragment 5 nm in size ("FAB-2") against the primary, or to a 10nm full length antibody against the primary ("Full length").
(b) Structures equivalent to the schematics described in (a) after angular averaging.
(c) Projection of the angularly averaged structure shown in (b) showing a double-peaked distribution

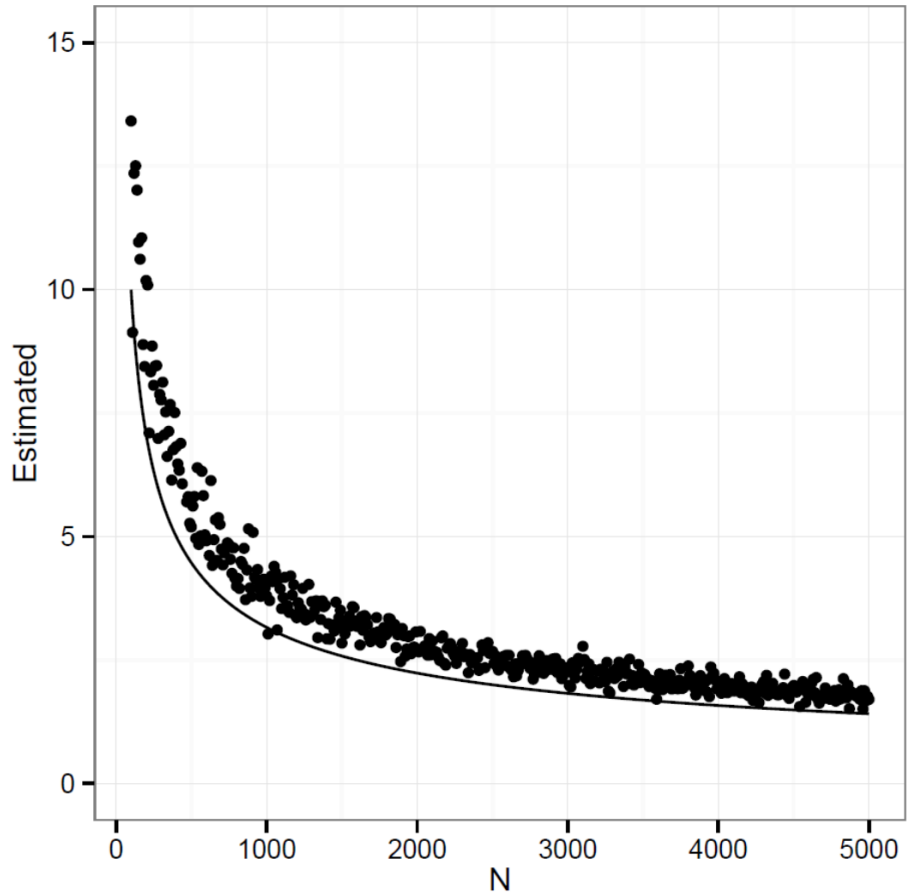

**Figure S4: Bootstrapping validation: How good is the bootstrapping method?**

Example of a Gaussian shape. Points are generated from a Gaussian shape with mean (µ) and sigma (σ). Points are binned and a Gaussian function fitted to the data, from which the µ is extracted. This fitted value of µ is bootstrapped to estimate its error. This plot shows the estimated error as a function of the number of points used in the fitting. The continuous line shows the CRB for the estimation of the µ parameter (sqrt(N)/N).

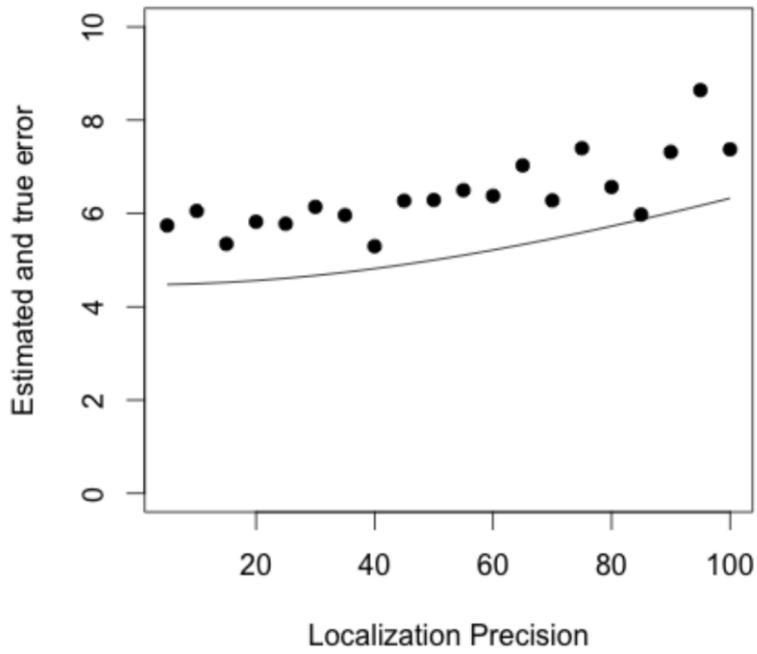

**Figure S5: Effects of localization precision on bootstrapped error**

Using the same example of a Gaussian shape (Figure S4), we estimate the bootstrapped error dependence on the localization precision.

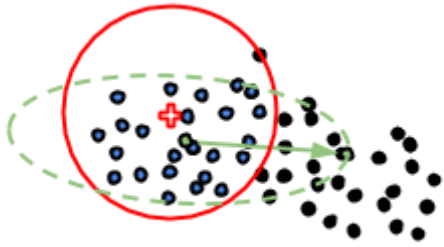

**Figure S6. Tracing algorithm.**

Starting from an initial position, the tracing algorithm follows the point cloud along the orientation within a specified search radius. Each step in the trace is the geometrical mean of all   points within the search radius.

Supplementary Methods

Cell Biology

African green monkey kidney cells (COS-7) were cultured in DMEM supplemented with 10% FBS (Sigma-Aldrich) in a cell culture incubator (37°C and 5% CO2). U2OS cells (European Collection for Cell Cultures) were maintained in McCoy's 5A GlutaMAX medium (Life Technologies) supplemented with 10% FBS in a cell culture incubator (37°C and 5% CO2). In both cases, cells were plated at low confluency on cleaned 25 mm #1 coverglass (Menzel). Coverglasses were cleaned by bathing overnight in 50% Glacial acetic acid (320099 Sigma Aldrich), 50% Ethanol, then rinsed in pure Ethanol and flamed.

Reactants Preparation

Mercaptoethylamine (MEA – Sigma-Aldrich 30070) was prepared as a 1 M stock solution in deionized water, then adjusted to ~pH 8 using glacial acetic acid (Sigma-Aldrich), and stored at 4°C. β-mercaptoethanol (BME - Sigma-Aldrich M6250) was stored undiluted (14.3 M) at 4°C. Cyclooctatetraene (COT – Sigma-Aldrich 138924) was reconstituted in pure DMSO as 200 mM stock solutions. PCA (Protocatechuic acid, Sigma-Aldrich 37580) was dissolved to 100 mM in deionized water and adjusted to pH 9 using KOH and stored at 4°C; PCD (Protocatechuic dioxygenase, Sigma-Aldrich P8279) was stored at −20°C in 50% glycerol in 50 mM KCl, 1 mM EDTA and 100 mM Tris-HCl pH 8 at a concentration of 5 µM. Pre-extraction buffer (BRB80 + 4mM EGTA + 0.5% Triton X-100 (Triton) with BRB80 = 80 mM PIPES (P8203, Sigma-Aldrich) 1 mM $MgCl_2$, 1 mM EGTA, adjusted to pH 6.8 with KOH, all Sigma-Aldrich) was prepared as a 5x stock solution, by solubilizing PIPES and EGTA with KOH 10M then adjusting the pH to 6.8 and adding $MgCl_2$. The 5x solution was stored at 4°C, and diluted in water before use.

Sample preparation for microtubules

Prior to fixation, PBS and pre-extraction buffer were pre-warmed to 37°C: 24 h after plating, Cos7 cells were rinsed in PBS, then pre-extracted for 20 s in the pre-extraction buffer, washed in PBS, fixed for 10 min in −20°C Methanol (Sigma-Aldrich), then washed again 3 times in PBS (Room temp, and from here on, all buffers at room temperature). The samples were then blocked for 30 minutes in 5% BSA, before being incubated for 1.5 h at room temperature with 1:1000 mouse alpha-tubulin antibodies (Sigma, T5168) in 1% BSA diluted in PBS −0.2% Triton (PBST), followed by 3 washes with PBST, and then incubated for 45 min in 1%BSA-PBST with 1:1000 goat anti-mouse Alexa Fluor 647 (Alexa-647) F(ab')2 secondary antibody fragments (Life Technologies, A-21237) followed by 3 more washed with PBST.

Alternatively, Alexa-647 Full length secondary antibody (Life technology A-21235) were used instead of Alexa-647 F(ab')2 secondary antibody fragments, or primary mouse anti-alpha-tubulin antibodies were directly conjugated using the APEX Alexa Fluor 647 antibody labeling kit (Life Technologies) according to the manufacturer's instructions.

Sample preparation for Cep152

U2OS cells fixation and immunostaining was performed similarly as for tubulin, except that the primary rabbit anti-Cep152 antibody (Sigma-Aldrich, HPA039408) was used at 1:2000 in 1% BSA - PBST, and the secondary antibody was goat anti-rabbit Alexa-647 F(ab')2 secondary antibody fragments (Life Technologies, A-21246) at 1:1000 in 1% BSA - PBST.

STORM imaging

STORM imaging was performed on a modified iX71 Olympus microscope were the objective turret is replaced by a drilled aluminum breadboard (Thorlabs) to increase stability. A piezo objective scanner (P-725 PIFOC, Physik Instrumente, Germany) is mounted on the breadboard, and a 100x/1.3NA oil objective (Olympus, UplanFL, Japan) is used. The sample was mounted on an XYZ piezo stage (PINANO, Physik Instrumente, Germany)) using a custom built sample holder. A 100 mW 641 nm laser (Coherent, CUBE 640-100C, USA) was transmitted through a telescope with varying divergence used to change the size of the beam in the focal volume, and then reflected by a multiband dichroic (89100 bs, Chroma,USA) on the back aperture of the objective. The collected fluorescence was filtered using a band-pass emission filter (ET700/75, Chroma) and imaged onto an EMCCD camera (IxonEM+, Andor) using a 1.6 magnifying lens and resulting in a 100 nm pixel. The EMCCD was used with the conventional CCD amplifier at a frame rate of 25 fps and with frame transfer activated and 10,000–20,000 frames were recorded and saved as 16bits tif files. Laser intensity on the sample measured after the objective was 30-50 mW, and the FWHM of the laser beam at the focus was 25 μm (measured with ImageJ on a laser reflection). Assuming a Gaussian excitation, the average power density within the center of the square excitation defined as ([-FWHM/2 FWHM/2],[-FWHM/2 FWHM/2]) can be approximated as:

$$P(kW.cm^{-2}) \approx \frac{57 * I(mW)}{FWHM(\mu m)^2}$$

So for I = 30 mW at the focus we get 2.5kW.cm$^{-2}$

STORM Buffer

STORM imaging of both microtubules and CEP 152 was performed in 10 mM PBS-Tris pH 7.5 with 10 mM MEA combined with 50 mM BME, 2 mM COT, 2.5 mM PCA and 50 nM PCD. 2ml of this buffer was prepared in a 2ml eppendorf tube, and added on top of the sample ~10 minutes prior to imaging. Air exchanges were limited by a piece of dark tape placed on top of the sample holder, but not completely blocked.

STORM Data Analysis

Peak fitting was performed with Peakselector (courtesy of H. Hess) Each peak with a high enough signal-to-noise ratio was fitted to a Gaussian function and analyzed, and photon counts were extracted from the fitted peaks, using the calibrated camera sensitivity. Outliers (peaks detected for more than 15 consecutive frames, and peaks with too high of a residual after fitting) as well as peaks localized with less than 1000 photons were removed from the analysis. Peaks detected in successive frames at a distance of less than 60 nm were grouped and considered as a single molecule. Grouping of successive localizations was performed using MATLAB. Localized peaks were tracked in 2D (x–y) using a single particle tracking algorithm (http://physics.georgetown.edu/matlab/index.html ) with a maximum search radius of 60 nm, and all the localizations in a track were averaged to give a final molecular location (and the associated standard deviation), as well as molecular number of photons.

FRC resolution (Figure S2) was computed with MATLAB using the code provided by the authors. The resolution was computed for non-overlapping sub-regions of the images, and outliers (regions for which the FRC resolution was larger than 200nm or failed to converge) were removed before averages and standard deviation were computed.

Tracing (figure 2 and 3; Figure S6) was performed with MATLAB using PALMsiever and the Tracer plugin. The tracing algorithm is given an initial position $P_0$ (red cross) a direction dP, a search radius *r* and a step size *s*. It first searches around the initial position for all points within the search radius (blue points); then, it calculates their orientation by calculating the principal components of the covariance matrix (green ellipse). It then sets the geometrical mean of the point cloud as the first point in the trace (green point) and moves along the strongest component (green arrow) for a distance *s*. It repeats this process with unvisited points until there are less than 20 unvisited points within the search radius.

BMC (figure 2, 3 and 4) was performed with MATLAB using PALMsiever and the BMCcircleFit and BMCtraceAnalyzer plugins (provided in the supplementary code). The BMCcircleFit algorithm collects all the (N) points within the current view. For each number of points N' in the sequence {N, N/2, N/4, N/8, …} it samples 100 times with repetition from the collection of N' points, repeats a circle fit and calculates the mean and standard deviation of the 100 estimations. Additionally, the resulting curve is fit to a power law. The circle fit is performed by first estimating the center of the circle by calculating the centroid, and then fitting a histogram of the distances from the center to a single Gaussian function of the form

$$Ae^{-(x-B)^2/(2C^2)}$$

The BMCtraceAnalyzer plugin works in a similar way, except that the points are collected from an existing trace and the estimated function is in this case a double Gaussian function of the form

$$Ae^{-((x-B-C/2)^2/D^2)} + Ae^{-((x-B+C/2)^2/D^2)}$$

```matlab
% BMC circle fit plugin
function BMCcircleFit(handles)

m = msgbox('Select around the object you want to analyze, then double-click
the rectangle to continue');
h1 = imrect(handles.axes1);
pos = wait(h1);
close(m);

if ~isempty(pos)
    setBounds(handles,[pos(1) pos(1)+pos(3) pos(2) pos(2)+pos(3)]);
end

subset = getSubset(handles);
X=getX(handles); X=X(subset);
Y=getY(handles); Y=Y(subset);
bins=getRes(handles);

[radius xm ym h bins ft] = estimateRadius(X,Y,bins);
N = numel(X);
Ns = round(exp(fliplr(log(N):-(log(2)):log(10))));
logger(['Estimating radius mean and stddev for N=' num2str(Ns) '...']);

for ni = 1:numel(Ns)
    n = Ns(ni);
    
    % Random pick with repetition
    I = 1+round(rand(size(X))*(N-1)); I = I(1:n);
    radiuses = bootstrp(100, @(x,y) estimateRadius(x,y,bins), X(I), Y(I));
    
    r(ni) = mean(radiuses); var_r(ni) = sum( (radiuses-radius).^2 );
    logger(['Estimated radius of ' num2str(r(ni)) ' +- '
num2str(.5*2.35*sqrt( var_r(ni))) ' with ' num2str(n) 'points'])
end

snr = radius./sqrt(var_r);
figure; loglog(Ns, snr, '+'); xlabel('# Points'); ylabel('SNR');

% Power law fit
ab = polyfit(log(Ns),log(snr),1); a = exp(ab(2)); b = ab(1);
hold; plot(Ns,exp(ab(2))*Ns.^ab(1),'r');
logger(['Power law fit: ' num2str(a) ' N^{' num2str(b) '}'])

% How many points for SNR of 3?
n_est_snr3 = round(exp(log(3/a)/b));
logger(['For an SNR of 3, an estimated ' num2str(n_est_snr3) ' points are
needed.']);

% How many points for resolution of 5nm?
figure; loglog(Ns, sqrt(var_r), '+'); xlabel('# Points'); ylabel('BMC of
the radius');
abs = polyfit(log(Ns),log(sqrt(var_r)),1); as = exp(abs(2)); bs = abs(1);
hold; plot(Ns,exp(abs(2))*Ns.^abs(1),'r');
n_est_res5 = round(exp(log(5/as)/bs));
logger(['Power law fit: ' num2str(as) ' N^{' num2str(bs) '}'])
logger(['For a resolution of 5, an estimated ' num2str(n_est_res5) ' points
are needed.']);
```

```matlab
% BMC trace analyzer
function BMCtraceAnalyzer(handles)
% Gather trace, x, y, subset from workspace
[Trace, subset, X, Y] = 
fetch('Trace','subset',handles.settings.varx,handles.settings.vary);
X=X(subset); Y=Y(subset);

nbins = str2double(get(handles.tBins,'String'));
r = str2double(get(handles.radius,'String'));

[A, centersY , ~, centersX] = trace_histogram(Trace, X, Y, r, nbins, 0);

[sX, sY]=trace_collect(Trace, X, Y, r, 1);

width = getfield(dg_fit(centersY',histc(sY,centersY)),'w');

N = numel(sY);
Ns = round(exp(fliplr(log(N):-(log(2)):log(50))));
centersY = linspace(-r,r,nbins);
logger(['Estimating width mean and stddev for N=' num2str(Ns) '...']);
for ni = 1:numel(Ns)
    n = Ns(ni);
    
    % Random pick with repetition
    I = 1+round(rand(size(sY))*(N-1)); I = I(1:n);
    widths = bootstrp(100, @(x) 
getfield(dg_fit(centersY',histc(x,centersY)),'w'),sY(I));
    
    w(ni) = mean(widths); var_w(ni) = var(widths);
    logger(['Estimated width ' num2str(w(ni)) ' +- ' num2str(.5*2.35*sqrt(
var_w(ni))) ' with ' num2str(n) 'points'])
end

logger(sprintf('RESOLUTION on width estimation : 
%f',2.35*sqrt(var_w(end))))

snr = width./sqrt(var_w);

figure; loglog(Ns, snr, '+'); xlabel('# Points'); ylabel('SNR');

% Power law fit
ab = polyfit(log(Ns),log(snr),1); a = exp(ab(2)); b = ab(1);
hold; plot(Ns,exp(ab(2))*Ns.^ab(1),'r');
logger(['Power law fit: ' num2str(a) ' N^{' num2str(b) '}'])

% How many points for SNR of 3?
n_est_snr3 = round(exp(log(3/a)/b));
logger(['For an SNR of 3, an estimated ' num2str(n_est_snr3) ' points are
needed.']);

% How many points for resolution of 5nm?
figure; loglog(Ns, sqrt(var_w), '+'); xlabel('# Points'); ylabel('BMC of 
w');
abs = polyfit(log(Ns),log(sqrt(var_w)),1); as = exp(abs(2)); bs = abs(1);
hold; plot(Ns,exp(abs(2))*Ns.^abs(1),'r');
n_est_res1 = round(exp(log(1/as)/bs));
logger(['Power law fit: ' num2str(as) ' N^{' num2str(bs) '}'])
logger(['For a resolution of 1, an estimated ' num2str(n_est_res1) ' points
are needed.']);
```